# Elastically Cooperative Activated Barrier Hopping Theory of Relaxation in Viscous Fluids. I. General Formulation and Application to Hard Sphere Fluids

Stephen Mirigian and Kenneth S. Schweizer*

Department of Materials Science, Department of Chemistry, and Frederick Seitz Materials Research Laboratory, University of Illinois, 1304 W. Green Street, Urbana, Illinois 61801

* kschweiz@illinois.edu

## Abstract

We generalize the force-level Nonlinear Langevin Equation theory of single particle hopping to include collective effects associated with long range elastic distortion of the liquid. The activated alpha relaxation event is of a mixed spatial character, involving two distinct, but inter-related, local and collective barriers. There are no divergences at volume fractions below jamming or temperatures above zero Kelvin. The ideas are first developed and implemented analytically and numerically in the context of hard sphere fluids. In an intermediate volume fraction crossover regime, the local cage process is dominant in a manner consistent with an apparent Arrhenius behavior. The super-Arrhenius collective barrier is more strongly dependent on volume fraction, dominates the highly viscous regime, and is well described by a nonsingular law below jamming. The increase of the collective barrier is determined by the amplitude of thermal density fluctuations, shear modulus or transient localization length, and a growing microscopic jump length. Alpha relaxation time calculations are in good agreement with recent experiments and simulations on dense fluids and suspensions of hard spheres. Comparisons of the theory with elastic models and entropy crisis ideas are explored. The present work provides a foundation for constructing a quasi-universal, fit-parameter-free theory for relaxation in thermal molecular liquids over 14 orders of magnitude in time.

## I. Introduction

### A. General Background

The cooling of glass-forming liquids leads to a spectacular increase of the structural relaxation time and viscosity by 14 or more orders of magnitude, the interpretation of which usually invokes physically distinct regimes separated by material-specific "crossover temperatures" [1-4]. As lucidly discussed by Tarjus [5], the many conceptually distinct theories [6-8] are built on different combinations of answers to the following questions. (i) Is there a finite temperature or below jamming divergence? (ii) Is slow dynamics intrinsically kinetic or causally tied to thermodynamics? (iii) Are postulated critical points avoided, unreachable or nonexistent? (iv) Is dynamic heterogeneity a second order generic consequence of activated motion or the driver of vitrification? (v) Is the alpha process described at a coarse grained order parameter level or as a specific collective motion? Confrontation of theories with experiment are usually carried out over a limited temperature range and involve multiple fit parameters, and few are formulated at the level of forces and molecules, which renders a definitive evaluation of competing ideas difficult. Simulations suggest existing "microscopic" theories have significant problems in capturing the collective aspects of activated processes [9].

The only first principles approach which relates forces, structure and slow relaxation is often stated to be ideal mode coupling theory (IMCT) based on collective pair density fluctuations [10, 11]. Its signature achievement is the prediction of particle localization and emergence of rigidity. Unfortunately, its dynamically Gaussian nature results in literal kinetic arrest far above $T_g$ due to neglect of activated processes. IMCT can empirically fit only a few orders of magnitude of growth of the alpha time in the dynamic "precursor" regime where there are many indications activated motions are already important, e.g., large nongaussian parameters, relaxation-diffusion decoupling, exponential tails of the van Hove function, and non-MCT wavevector dependence of the relaxation time beyond the cage scale [6,8,12-14]. Some have suggested [14] IMCT is, at best, applicable to liquids for a small "sliver" of intermediate temperatures, and *never* for the alpha relaxation time.

The above strengths and weaknesses of IMCT motivated the development of the nonlinear Langevin equation (NLE) approach [15, 16] which builds on a simplified "naïve" MCT (NMCT) [17] framework to address *single* particle hopping in a dynamic mean field spirit (Fig. 1). Barriers are finite below jamming and at nonzero temperature, and the unphysical MCT singularity is removed. Hard sphere relaxation in the precursor regime and hopping-induced nongaussian phenomena due to trajectory fluctuations are rather well described over the limited volume fractions accessible in colloidal suspensions [12, 13, 18]. NLE theory predicts an *apparent* critical power law scaling of the alpha time over ~3 orders of magnitude as a consequence of hopping over low entropic barriers [15]. However, NLE theory is a local cage approach which does not explicitly capture longer range collective motions, a presumably fatal limitation in deeply supercooled liquids or ultra-concentrated suspensions as suggested by recent simulations [9] and theoretical analysis [19].

The goal of the present paper is to qualitatively extend NLE theory to explicitly include collective effects. This first article focuses on hard sphere fluids, and in the companion paper II we propose and implement a mapping of thermal liquids to an effective hard sphere fluid. The article is long since we aim for a detailed motivation and description of our ideas, supported by analytic and numerical analyses, and comparison to experiment and simulation. We also contrast our approach with other models and theories. A brief Letter reporting the new approach was recently published [20].

**B. Thermal Liquid Approaches and Phenomenology**

For the question of the emergence of localization and material rigidity, MCT [10], NLE [16] and replica theory (RT) [8,21] are all in qualitative accord: a crossover temperature ($T_A$) or volume fraction ($\phi_A$) is predicted beyond which relaxation requires surmounting barriers. In potential energy landscape (PEL) simulations this is called the landscape onset transition [22], and it occurs *well above* the *empirically-deduced* MCT transition temperature $T_c$. How *finite* barriers are theoretically computed, and the mechanism for restoring ergodicity via hopping, remains a difficult and controversial problem. Replica theory [8,21], a literal mean field thermodynamic approach, predicts below $T_A$ barriers are of infinite height, and at a lower Kauzmann temperature $T_K$ (well below $T_g$) a phase transition occurs where the configurational entropy or complexity vanishes. To address this gross over-prediction of barriers, the random first order phase transition theory (RFOT) [23] adds a non-perturbative decoration of droplets to RT. In a nucleation spirit, RFOT is based on a compact mosaic size as the relevant order parameter, which must reach a critical radius to "unlock" the frozen localized state on smaller scales predicted by the underlying RT or MCT description [23-25]. The diversity of packing structures drives structural relaxation which is resisted by a nonclassical surface tension resulting in a barrier that scales inversely with configurational entropy. No direct treatment of particle motions enter except the qualitative notion that all particles in the mosaic displace a distance of order the localization or Lindeman length.

There are two fundamental differences between our new approach and any theory built on a literal mean field (RT or MCT) perspective. First, our starting point is neither a thermodynamic nor kinetic "ideal glass" picture. Rather, the unphysical MCT divergence is first removed using the NLE idea which serves as the foundation for subsequent development. The theoretical problem then becomes the opposite, i.e., due to its single particle dynamic mean field nature, NLE theory *under*-predicts, not over-predicts, barriers. Second, although slow relaxation is related to equilibrium properties, connections to thermodynamics are a "consequence" of the force-level treatment.

To generalize the NLE approach to include collective effects, we are strongly motivated by the *smooth* pattern of growth of the relaxation time in fragile thermal liquids over 14-15 orders of magnitude: *effectively* Arrhenius at high temperature (barrier, $E_A$) and strongly super-Arrhenius at low temperature, separated by a relatively narrow "curved" intermediate or crossover regime [1,3,4,26-28]. We believe the most straightforward physical interpretation of this pattern is a "two-barrier" picture of the alpha process, in the spirit of the phenomenological work of Kivelson and Tarjus [29-31] and others in polymer science [32, 33] and solid-state physics [34]. Very recent experiments of Rossler and coworkers [27, 28, 35] have convincingly established a phenomenological two-barrier picture works extremely well over a remarkably wide temperature range covering alpha times ranging from psecs to >100 s; moreover, at $T_g$ the two barriers are related in a nearly universal manner for van der Waals liquids. The deduced high temperature activated regime is consistent with Roland's analysis [3, 4], Goldstein's arguments [36], and simulations of liquid metals and Lennard-Jones (LJ) model fluids [9,37-40]. More generally, the existence of empirical correlations between $E_A$, super-Arrhenius relaxation, $T_g$ and fragility [26-28,35,41] suggests the underlying physics that determines each barrier is not independent.

The dynamic crossover temperature signals the beginning of the deeply supercooled regime. In PEL simulations [22] it is associated with barriers of significant height that at the *empirical* MCT $T_c$ are already large, ~10 $k_BT$, corresponding to an alpha time ~10-100 ns [8]. This is consistent with many simulations of LJ and related atomic models that find (in reduced units) $T_A \sim 1$, far above the empirical $T_c \sim 0.45$, and in good agreement with *a priori* MCT calculations. [22, 42]. This suggests to us (consistent with NLE studies [13]) that the *predicted* MCT critical temperature corresponds to $T_A$, and the empirically-deduced dynamic crossover "$T_c$" indicates a change in the nature of the activated process. If one accepts this picture, as we do, then activated dynamics plays a crucial role over the entire range of temperatures associated with supercooled liquids. The question then becomes: What is the detailed physical nature of the empirically-deduced dynamic crossover?

### C. Collective Elastic Effects in Viscous Liquids and Colloidal Suspensions

We will show that the local cage NLE approach provides a microscopic basis for a non-cooperative Arrhenius-like activated process in both hard sphere and thermal liquids, but is dominant only in a rather narrow intermediate crossover regime. At high temperatures (lower volume fractions for hard spheres) it is partly obscured by (dressed) binary collision physics. At low temperatures (high volume fractions), a longer range collective effect dominates that is absent in NLE theory. Our working hypothesis, partially motivated by the diverse experimental, phenomenological modeling, and simulations studies discussed below, is the key collective physics is emergent elasticity.

A fundamental connection between barriers and small scale elastic/vibrational-like motion (top and bottom of the PEL) is broadly supported by the many striking correlations between fast dynamics (characteristic of the transiently localized state) and the slow alpha process [1, 3, 26, 41, 43]. It also finds empirical support in the ability of divergence-free phenomenological "elastic models" [44-53] (e.g., shoving [46]), inspired by the idea that highly viscous liquids are solids that flow, to describe relaxation in the deeply supercooled regime. Even Goldstein [36] argued for a strong elastic physics component of the alpha process including central ideas underlying Dyre's shoving model, namely: (a) cage expansion is required to allow an activated local molecular rearrangement to occur, (b) the crucial resisting property is the temperature-dependent (relaxed plateau) shear modulus, and (c) a long range, low amplitude strain field surrounds the local activation event. The key idea is the elementary relaxation event has a mixed local/nonlocal spatial character. Such a view seems to conflict with the Adams-Gibbs cooperatively-re-arranging region (CRR) [54], entropic droplet (RFOT) [23], or shear transformation zone (STZ) [55] ideas based on compact domains.

We believe multiple experiments support the viewpoint that the elementary relaxation event is of this mixed local/nonlocal character. Confocal imaging of glassy colloidal suspensions finds the irreversible dynamical event consists of a compact core of large amplitude particle displacements surrounded by a long range elastic dressing; the long range particle displacements play a major role in determining the entropic barrier and are well described by continuum elasticity theory [56]. This early study has been buttressed in recent experiments of Schall and coworkers who have established the presence of long range strain correlations associated with small amplitude colloid displacements and hence the nonlocal nature of the alpha process [57, 58]. Most recently, Weeks and coworkers [59] used local force probe measurements in hard sphere suspensions and found a long range strain field of an isotropic homogeneous elastic medium form surrounds an induced probe displacement. Interestingly, the same qualitative physical picture proposed for colloids [56] was indirectly deduced far earlier to apply to molecular liquids based on NMR measurements [60].

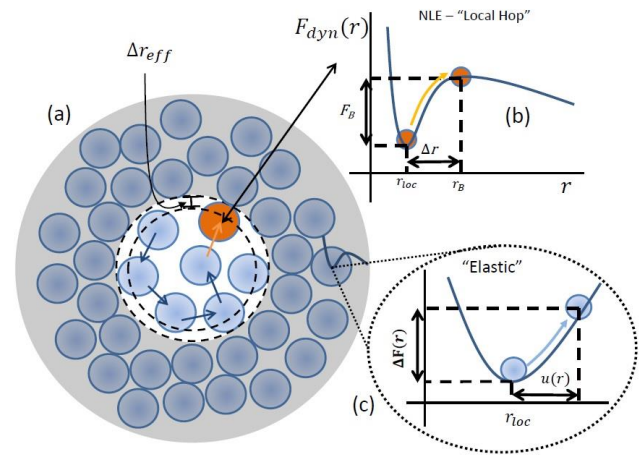

**Fig 1:** (a) Schematic of the local hopping process described by NLE theory leading to a long ranged collective elastic barrier. (b) Rearrangement on the length scale of the local cage is described by the NLE dynamic free energy barrier. (c) The collective elastic barrier arises from the small harmonic displacements of all particles outside the cage.

Multiple recent simulation studies have established the key role played by long range strain/stress fields, emergent elasticity in the liquid state, and the importance of activated events even at "high" temperature. Abraham and Harrowell [61] found stress relaxation changes from its simple un-activated high temperature behavior to a "low" temperature inherent structure dominated process well above the empirical $T_c$ at a temperature close to both $T_A$ and what is expected from the elasticity-based crossover ideas of Trachenko et. al. [62, 63]. Egami and coworkers have simulated liquid metals and found: (i) apparent Arrhenius relaxation at high temperatures, (ii) emergent elastic distortions (apparent strains of order a few percent on the cage scale) associated with local configurational re-arrangements, and (iii) a remarkably long range character (10 particle diameters or beyond) of the stress correlations that determine the viscosity thereby establishing it as a fundamentally nonlocal quantity [40, 64, 65]. Most recently, Chattoraj and Lemaitre [66] have shown that even in the supercooled liquid regime the activated irreversible local relaxation event is associated with a long range deformation field, as previously seen in low temperature glasses. They conclude this provides strong support for the idea that supercooled liquids are "solids that flow" [49].

### D. Our Approach

Fig. 1a sketches the physical essence of our approach [20]. The central concept, phenomenologically embedded in elastic models, is that when a liquid is sufficiently cold or dense, for steric reasons the local cage scale hopping event requires increasing participation by the surrounding medium in the form of long range particle displacements resulting in an additional barrier. The total activation event is then of a mixed local/nonlocal character involving two distinct, *but intimately related*, barriers: one due to the local cage scale rearrangement (described by NLE theory) and one due to the elastic-like motion of the surrounding fluid necessary to accommodate this local rearrangement. These two contributions determine a single total barrier that controls the fundamental relaxation event. Our analysis qualitatively extends elastic models to include a local hopping process, how it microscopically couples to elasticity, *and* a growing length scale. The empirically-deduced dynamical crossover then corresponds not to an avoided MCT transition, but rather to when the amplitude of the local activated event is sufficiently large that the elastic energy cost becomes more important than the local barrier in determining the rate of growth of the alpha time. No critical points (literal or avoided) or divergences of any kind enter below jamming or above zero Kelvin.

Section II recalls the key elements of NLE theory. Its extension to include elastic effects is derived in section III, and an analytic analysis of the total barrier, including the jamming limit, is presented. Numerical calculations of the alpha time are given in section IV, and compared to diverse phenomenological models and experiments/simulations on hard sphere fluids. Comparison of our approach to elastic and entropy crisis models is the subject of section V. Nonexponential relaxation and growing length scales are discussed in section VI. The paper concludes with a summary in section VII. The companion paper II proposes a no adjustable parameter mapping of thermal liquids to a reference hard sphere fluid, and performs quantitative calculations and comparisons to experiment and other theories.

## II. Normal Fluid and Nonlinear Langevin Equation Theories

We first briefly recall the well-established theories for the normal liquid and non-cooperative hopping regime that provide the starting point for our development.

### A. Non-Activated Theory

For hard spheres (diameter, d), the "binary collision mean field" (BCMF) [67-69] theory captures the Enskog regime and the mean (non-self-consistent) effect of local caging in normal fluids. It describes simulations and colloid experiments (diffusion constant, viscosity, relaxation times) well up to a volume fraction $\phi = \pi \rho d^3 / 6 \sim 0.5$ where transport coefficients are modified by a factor of ~ 20-30 relative to the dilute limit. The BCMF results enter NLE theory via the short time diffusion constant, $D_s$, or time scale $\tau_s$, and we adopt the "s" subscript notation to write down its formulas:

$$D_s = \frac{k_B T}{\zeta_s} \quad (1)$$

$$\zeta_s = \zeta_0 \left[ 1 + \frac{d^3}{36\pi\phi} \int_0^\infty dk \, k^2 \frac{(S(k)-1)^2}{S(k)+b(k)} \right] \quad (2)$$

$$b^{-1}(k) \equiv 1 - j_0(kd) + 2 j_2(kd) \quad (3)$$

where $j_n(x)$ is the spherical Bessel function of order n. The "bare" friction constant for a colloidal suspension is $\zeta_0 = \zeta_{SE} g(d)$ where g(d) is the contact value of the pair correlation function and $\zeta_{SE}$ is the Stokes-Einstein friction constant, and for a Newtonian fluid is the Enskog friction constant $\zeta_0 = \zeta_E$. A short relaxation time scale is defined as:

$$\tau_s \equiv \frac{d^2}{D_s} = \tau_0 \left[ 1 + \frac{1}{36\pi\phi} \int_0^\infty dQ \, \frac{Q^2 (S(Q)-1)^2}{S(Q)+b(Q)} \right] \quad (4)$$

where Q=kd, the bare time for Newtonian fluids is [70]

$$\tau_0 = g(d)^2 \tau_E = \frac{g(d)}{24\rho d^2} \sqrt{\frac{M}{\pi k_B T}} \quad (5a)$$

and $\rho$ is the particle number density. For suspensions the bare time is [69]:

$$\tau_0 = g(d) \frac{\zeta_{SE} d^2}{k_B T} = \frac{1+\phi/2}{(1-\phi)^2} \cdot \frac{\zeta_{SE} d^2}{k_B T} \quad (5b)$$

where the final equality uses the Percus-Yevick (PY) value [70] for the contact value.

### B. Single Particle Barrier Hopping

NLE theory has been extensively developed over the last decade primarily to treat diverse colloidal suspensions (glassy, gels) in the dynamical precursor, but activated, regime [69-75]. For spheres, it is based on an overdamped equation-of-motion for the scalar (angularly averaged) displacement of a tagged particle from its initial position, $r(t)$,

$$0 = -\zeta_s \frac{dr(t)}{dt} - \frac{\partial F_{dyn}(r(t))}{\partial r(t)} + \delta f(t) \quad (6)$$

where the thermal noise term satisfies $\langle \delta f(0)\delta f(t)\rangle = 2k_BT\zeta_s\delta(t)$, and $r(t=0)=0$. The key quantity is the *dynamic* free energy, $F_{dyn}(r)$, which describes the effective force on a moving tagged particle due to the surrounding fluid (number density $\rho$), and is given by:

$$\beta F_{dyn}(r) = -3\ln(r) + \int \frac{d\vec{k}}{(2\pi)^3}\frac{\rho C^2(k)S(k)}{1+S^{-1}(k)}\exp\left[-\frac{k^2r^2}{6}(1+S^{-1}(k))\right] \quad (7)$$

where $\beta \equiv (k_BT)^{-1}$, and C(k) and S(k) are the direct correlation function and structure factor, respectively.

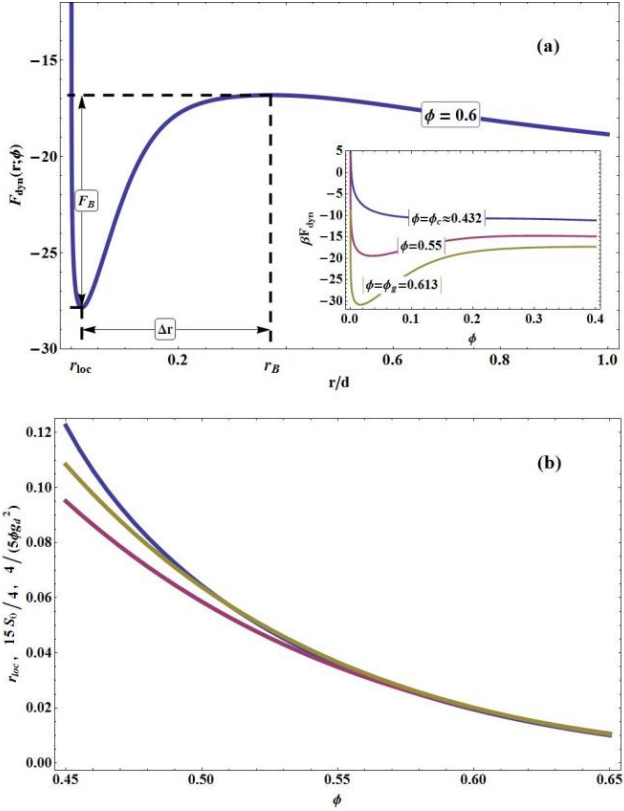

**Fig. 2:** (a) Dynamic free energy with relevant length and energy scales marked; Inset shows how it changes with increasing volume fraction. (b) From top to bottom (at low $\phi$): Localization length (blue), $\sim 1/\phi g_d^2$ (yellow), and dimensionless compressibility (red), showing their proportionality at higher volume fractions.

Fig. 2a shows an example dynamic free energy with relevant length and energy scales labeled, and how it changes with volume fraction. Based on PY theory input, the competition between the localizing and delocalizing contributions in Eq.(7) results in the emergence of a local minimum of $F_{dyn}(r)$ at $\phi \equiv \phi_A \approx 0.43$, signaling the onset of *transient* localization and a barrier. The two important length scales in $F_{dyn}(r)$ are the localization length $r_{loc}$ and barrier location, $r_B$. One sees from Fig.2a the former decreases significantly with $\phi$, while the latter weakly increases; the "jump distance", $\Delta r \equiv r_B - r_{loc}$, increases as barriers grow. Three key energy scales are the well (barrier) curvature $K_0$ ($K_B$) in units of $k_BT/d^2$, and barrier height, $F_B(\phi)$, in $k_BT$ units.

If the noise in Eq.(6) is dropped, an ideal glass transition of the NMCT type is predicted based on the long time mean square displacement or localization length, $\langle r^2(t\to\infty)\rangle \equiv r_{loc}^2$, which satisfies the self-consistent relation [16]:

$$\frac{1}{2}\beta\langle \vec{f}(0)\bullet\vec{f}(t\to\infty)\rangle r_{loc}^2 = \frac{3}{2}k_BT \quad (8)$$

where $\vec{f}(t)$ is the total force on a tagged particle at time t. Explicitly, Eq.(8) is

$$\frac{1}{r_{loc}^2} = \frac{1}{9}\int_0^\infty \frac{dk}{(2\pi)^3}V(k;\phi)e^{-\frac{k^2r_{loc}^2}{6}}e^{-\frac{k^2r_{loc}^2}{6S(k)}} \quad (9)$$

where $V(k)$ is a dynamical vertex that quantifies the effective kinetic constraints on a length scale $2\pi/k$ and is determined by liquid pair structure as

$$V(k;\phi) = 24\frac{\phi}{d^3}k^4C^2(k)S(k) \quad (10)$$

For hard spheres the "ultra-local" nature [76, 77] of dynamical caging in NLE theory was established analytically in the sense the dominant vertex contributions are from large wavevectors, $k > k_c = 2\pi/r_{cage}$, where $r_{cage} \approx (1.3-1.5)d$ is the location of the first minimum of g(r).

Kramers mean first passage time for barrier crossing is [78, 79]:

$$\frac{\tau_{hop}}{\tau_s} = \frac{2\pi}{\sqrt{K_0K_B}}\exp(\beta F_B) \quad (11)$$

When barriers are low, a mean first passage time can be computed as the average time for a particle to move from the localization length to the top of the barrier:

$$\frac{\tau_{hop}}{\tau_s} = d^{-2}\int_{r_{loc}}^{r_B} dx\, e^{\beta F_{dyn}(x)}\int_{r_{loc}}^x dy\, e^{-\beta F_{dyn}(y)} \quad (12)$$

The alpha time measured in diverse experiments and simulations show nearly identical behaviors [1]. Whether a collective or single particle relaxation time is measured seems to matter little with regard to its thermodynamic state dependence and apparently even absolute magnitude [28,35]. Hence, we compute it in a simple, generic manner as [80]:

$$\tau_\alpha \equiv \tau_s + \tau_{hop} \quad (13)$$

This is consistent with the short time friction in Eq.(6) which sets the time scale for hopping and the applicability of NLE theory. Although a barrier emerges at $\phi_A \sim 0.432$, it is subdominant until $\tau_s = \tau_{hop}$, which we find corresponds to $F_B \sim 3\ k_BT$ at $\phi = \phi_X \sim 0.53$.

The glassy shear modulus can be calculated as [81, 82]:

$$G = \frac{k_B T}{60\pi^2} \int_0^\infty dk \left[ k^2 \frac{d}{dk} \ln(S(k)) \right]^2 \exp\left(-\frac{k^2 r_{loc}^2}{3 S(k)}\right) \quad (14)$$

In reality, hopping restores ergodicity at long time scales, but Eq.(14) is an appropriate measure of rigidity on a time scale before activated events occur. Analytic analysis when $F_B$ is beyond several $k_BT$ has allowed the derivation of many connections between dimensionless features of the dynamic free energy [76]:

$$d^2 K_0 = \frac{3d^2}{r_{loc}^2} \propto \frac{Gd^3}{k_B T} \propto (\beta F_B)^2 \propto \nu_\infty^2 \quad (15)$$

Here, $\nu_\infty = 96\pi^2 \phi g^2(d)$ is a "coupling constant" that quantifies the strength of the effective mean square force caging a particle; it diverges as a double pole as jamming is approached [76], $\nu_\infty \propto (\phi_{rcp} - \phi)^{-2}$. Embedded in Eq.(15) are connections between the short and long time dynamics, elasticity, alpha time and barrier, which will play a key role in our predictions for the collective barrier. Fig. 2b shows the analytic result, $r_{loc} \approx 24\pi\sqrt{3\pi}\nu_\infty^{-1}$, works well for barriers in excess of a few $k_B T$. This local barrier has a simple physical interpretation since it is proportional to the maximum force confining a particle in its cage, $f_{max} \propto k_B T / r_{loc}$.

Kinetic-thermodynamic connections have been *numerically* established via the dimensionless compressibility, $S(k=0) \equiv S_0 = \rho k_B T \kappa_T$ ($B = 1/\kappa_T$ is the bulk modulus), which quantifies the amplitude of "long wavelength" (typically beyond ~ nm) density fluctuations. Beyond $\phi_X$ we find $r_{loc} \propto S_0$ (Fig.2b), and thus from Eq.(15) the local barrier scales linearly with the bulk modulus. Such a relationship was deduced long ago by Sjogren [83] in a different manner based on "extended-MCT".

Recent analyses of NLE theory for hard spheres with sophisticated structural input that includes jamming physics [19, 84] suggest it captures rather well roughly the first 3 decades of slow relaxation and many nongaussian fluctuation effects, up to $\phi \approx 0.57 - 0.58$ where $F_B \sim 7-8\ k_BT$. However, at higher $\phi$ the theory increasingly under predicts the alpha time and is not nearly fragile enough [19, 84]. This establishes NLE theory as useful for the crossover regime where local hopping is dominant and its breakdown in the ultra-dense regime where presumably collective processes dominate.

## III. Elastically Collective NLE Theory

We now extend NLE theory to include collective effects inspired by the elastic shoving model perspective that viscous liquids behave as "solids that flow"[48]. We call this approach the "elastically collective nonlinear Langevin equation" theory (ECNLE).

### A. Formulation

A core physical idea of NLE theory is that the alpha relaxation is associated relatively large amplitude ($\Delta r \approx 0.25d - 0.4d$) hopping of particles on the cage scale. However, its single particle dynamical mean field nature is expected to fail at sufficiently high volume fractions since such large amplitude motions cannot be accommodated without creating space in the increasingly rigid surroundings. Thus, the activation event and net barrier consists of two contributions: a core region of diameter $2r_{cage} \approx 3d$ where particles undergo non-vibrational hopping displacements, plus a strain field that dresses the local excitation (Fig.1). Outside the cage, a harmonic Einstein glass picture is adopted consistent with (see below) the small amplitude of elastic distortion. This physical picture is supported by colloidal microscopy observations [56] of rearrangement events in hard sphere suspensions which consist of a core of radius $\approx 1.5d$ dressed by an elastic distortion field, where particles in the core move $\sim 0.5 \pm 0.25 \mu m \approx 0.3d$.

To compute the collective elastic barrier requires an approximation consistent with the scalar (isotropic) and dynamic mean field nature of NLE theory. The former aspect motivates pre-averaging the angular part of the particle displacements of size $\Delta r = r_B - r_{loc}$ in the cage, resulting in an average *outward* radial motion (from the cage center) of $\frac{1}{2} \cdot \frac{2\pi}{4\pi} \int_0^{\pi/2} d\theta \Delta r \sin\theta = \frac{1}{4} \Delta r \equiv x_{rad}$, where the factor of ½ is due to half the particles moving inwards. Starting from an initial random position in the cage, $z$, the hopping motion perturbs the surroundings to a distance of $z + x_{rad} - r_{cage}$. If $z < r_{cage} - x_{rad}$, then outward particle motion occurs entirely within the cage and does not perturb the surroundings. Hence, the average cage expansion length scale is:

$$\Delta r_{eff} \equiv \frac{\int_{r_{cage}-x_{rad}}^{r_{cage}} dz(z+x_{rad}-r_{cage})z^2}{\int_0^{r_{cage}} dz\, z^2} \quad (16)$$

$$= \frac{3}{r_{cage}^3}\left[\frac{1}{3}\left(\frac{\Delta r}{4}-r_{cage}\right)\left(r_{cage}^3-\left(r_{cage}-\frac{\Delta r}{4}\right)^3\right)+\frac{1}{4}\left(r_{cage}^4-\left(r_{cage}-\frac{\Delta r}{4}\right)^4\right)\right]$$

$$= \frac{3}{r_{cage}^3}\left[\frac{r_{cage}^2 \Delta r^2}{32}+\frac{r_{cage}\Delta r^3}{192}-\frac{\Delta r^4}{3072}\right]$$

Using typical upper bound values of $\Delta r \approx 0.4d$ and $r_{cage} \approx 1.5d$ yields a very small degree of elastic displacement $\Delta r \ll r_{cage}$. We then obtain from Eq.(16):

$$\Delta r_{eff} \rightarrow \frac{3}{32}\frac{\Delta r^2}{r_{cage}} = \frac{3}{32}\frac{(r_B-r_{loc})^2}{r_{cage}}, \quad (17)$$

Cage expansion is thus orders of magnitude smaller than the hopping distance, $\Delta r_{eff} \ll \Delta r$. We also find it is of the order of, or smaller than, the transient localization length implying a harmonic description of the elastic distortion field is valid. Fig. 3 shows how the hopping distance and cage expansion displacement increase with volume fraction.

The above analysis of $\Delta r_{eff}$ is of course simplified. However, we believe it is logically consistent with the mean field character of isotropic NLE theory. An alternative calculation of $\Delta r_{eff}$ is presented in Appendix A which again yields $\Delta r_{eff} \propto \Delta r^2/r_{cage}$, but with a different numerical prefactor. For the rest of this article we adopt Eq.(16).

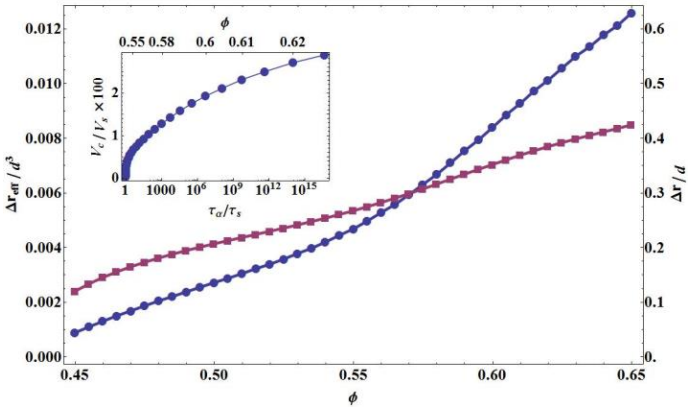

**Fig. 3:** Left axis shows the cage expansion (blue circles), right axis shows the hopping length scale (red squares) as a function of volume fraction. Inset: Growth in the cooperative volume versus $\log(\tau_\alpha/\tau_s)$; upper axis shows the volume fraction.

Since cage expansion is small, we adopt a continuum displacement strain field, ala the "shoving model" [85], which decays as $r^{-2}$. Imposing the boundary condition that the initial amplitude of the expansion is $\Delta r_{eff}$ yields [86]

$$u(r) = \Delta r_{eff}\left(\frac{r_{cage}}{r}\right)^2, \quad r > r_{cage} \quad (18)$$

where $r$ is the distance from the cage center. The dynamic free energy cost a distance $r$ from the cage center is thus

$$\Delta F(r) = F_{dyn}[r_{loc}+u(r)] - F_{dyn}[r_{loc}] \approx \frac{1}{2}K_0 u^2(r) \quad (19)$$

where the final result uses a harmonic approximation. The total elastic dynamic free energy cost follows by summing the contributions of all particles outside the cage region

$$\beta F_{elastic} = 4\pi \int_{r_{cage}}^{\infty} dr\, r^2 \rho g(r) \Delta F(r) \approx 12\phi \Delta r_{eff}^2 \left(\frac{r_{cage}}{d}\right)^3 K_0 \quad (20)$$

Since the strain field is long range, the relevant length scales are mainly when liquid structure is random, and hence $g(r)=1$ has been used; numerical calculations show this approximation incurs essentially negligible error. From Eq.(15), $K_0 \propto k_B T/r_{loc}^2$, providing a direct link between the localized state and collective barrier. Because the integrand of Eq.(20) decays as $r^{-2}$, the barrier attains 90% of its total value at a large distance, of order $\approx 10 r_{cage} \approx (13-15)d$, and in this sense the elastic barrier reflects long range physics. We shall demonstrate in the next section the collective barrier can become large compared to the local contribution. However, the mean square displacement (MSD) associated with the total excitation is modest since $\Delta r_{eff} \ll \Delta r < d$. More quantitatively, for a local cage of ~12 particles that move $\sim \Delta r \leq d/3$, the total cage scale MSD $\sim d^2$.

**B. Local and Collective Dynamic Barriers**

The total activation barrier is taken to be the sum of that due to the local cage rearrangement and the collective barrier:

$$F_{total} = F_B(\phi) + F_{elastic}(\phi) \quad (21)$$

Besides being the simplest thing to do, the barriers are added since we view the alpha process as a single event where in order for the local hop to be possible an expansive elastic cage fluctuation is required to create "free volume". This approach treats the two barriers on equal footing with no causal relationship implied. Specifically, the idea a cage scale hopping event "shoves" the surroundings is not necessary to invoke. Rather, an equally valid, and we feel more natural, physical interpretation is that an elastic fluctuation first occurs, thereby allowing the local rearrangement to occur.

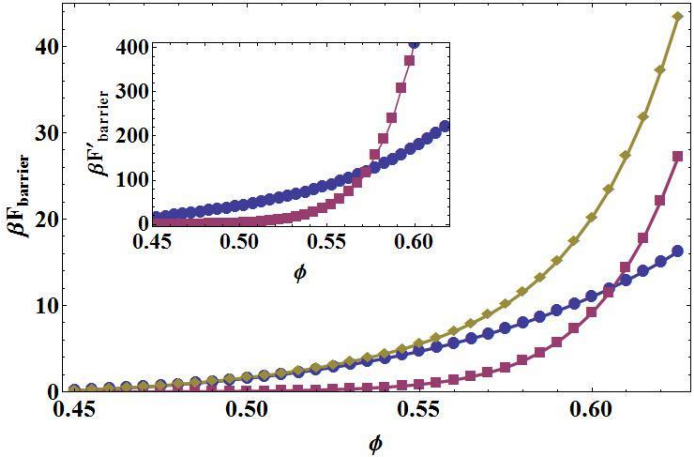

**Fig 4:** Different contributions to the total barrier: blue circles – NLE local barrier, red squares – collective elastic barrier, yellow diamonds – total barrier. Inset shows the volume fraction derivative of the local barrier (blue circles), and the collective elastic barrier (red squares).

Barrier height calculations are presented in the main frame of Fig. 4, while the inset shows volume fraction derivatives to quantify their different rates of *growth* with compaction. At lower volume fractions, both the magnitude and growth rate of the total barrier is dominated by the local barrier, $F_B$. The collective barrier is essentially negligible up to $\phi \approx 0.55$ (where it is $\sim k_B T$), and begins to increase faster than the local barrier at $\phi' \approx 0.57-0.58$ (see inset) thereby defining a crossover volume fraction, $\phi'$, which is curiously close to MCT empirical estimates of $\phi_c$. This smooth crossover occurs when $F_{total} \sim 10 k_B T$, corresponding to $\tau_\alpha \approx 10^4 \tau_s \approx 10^{-8}$ s in thermal liquids based on $\tau_s \approx p$s. At higher volume fractions the collective barrier grows rapidly and dominates the total barrier growth rate and dynamic fragility. However, its magnitude does not exceed the local barrier until close to kinetic arrest which occurs at $\phi \sim 0.61$ when $F_{tot} \sim 30 k_B T$.

The two barriers involve physically distinct motions, but are *not* independent, and are related to fluid structure and thermodynamics via $r_{loc} \propto S_0$ and Eq.(15). In a hypothetical (see below) ultra-viscous limit where $\Delta r_{eff}$ is nearly constant compared to the $\phi$-dependent changes of $K_0$ or the localization length, the predicted connection to *leading order* between the two barriers is remarkably simple and follows from Eq.(15) as [76]:

$$F_{elastic} \propto K_0 \propto G \propto S_0^{-2} \propto d^2/r_{loc}^2 \propto F_B^2 \qquad (22)$$

As discussed below, under experimental conditions there are corrections to $F_{elastic} \propto G \propto F_B^2$, as shown in Fig.5, and Eq.(22) applies only when $F_{total} \gtrsim 10 k_B T$, where

$$F_{total} \propto F_B(1+bF_B)\ , \ b \approx 0.15, \qquad (23)$$

### C. Analytic Analysis of the Barrier

To gain further insight, we now express the total barrier as simply as possible using the NLE theory ultra-local analysis results valid for high barriers. The proportionality constants in Eq.(15) are known, and one can write the local barrier as [76]

$$F_B \approx A \frac{r_{cage}}{r_{loc}} \frac{3\sqrt{3}\pi}{2\pi^2} \qquad (24)$$

The literal ultra-local analysis leads to Eq. (24) with A=1, but comparison with full numerical calculations suggest A=0.45 provides the most accurate analytic description. The cage radius varies little with volume fraction, from ~ 1.3-1.5 d. From Eq.(20), and using the ultra-local NLE analysis connections, the collective elastic barrier is

$$F_{elastic} = \frac{81 \phi \Delta r^4 r_{cage}}{256 r_{loc}^2} \qquad (25)$$

Thus, the full barrier can be written as

$$F_{total} \approx \frac{r_{cage}}{r_{loc}}\left(A\frac{3\sqrt{3}\pi}{2\pi^2} + \frac{81}{256}\phi\frac{\Delta r^4}{r_{loc}}\right) \qquad (26)$$

where energy and lengths are non-dimensionalized by $k_B T$ and $d$, respectively, and $\Delta r = r_B - r_{loc}$. The inset of Fig.5 shows Eq.(26) is surprisingly accurate. Prior analysis found the barrier location is a weakly (logarithmically) increasing function of volume fraction and can be analytically as approximated as [76]:

$$r_B \approx \frac{r_{cage}}{2\pi}\sqrt{3\ln\left(\frac{v_\infty}{24\pi^2}\right)} = \frac{r_{cage}}{2\pi}\sqrt{3\ln\left(\sqrt{\frac{3}{\pi}}\frac{1}{r_{loc}}\right)} \qquad (27)$$

Thus the *entire* barrier can be expressed in terms of the localization length. Given Eq.(27), one could replace (in a proportionality sense) the inverse localization length by $\sqrt{K_0}$, $\sqrt{G'}$, or $v_\infty$; these are all degenerate representations of the same physical ideas underlying ECNLE theory. The total barrier can thus be written to a leading order as the nearly universal expression

$$F_{total} \approx A\frac{3\sqrt{3}\pi}{2\pi^2}\frac{r_{cage}}{r_{loc}} + \frac{81}{256}\phi\frac{r_{cage}^5}{r_{loc}^2}\left(\sqrt{\frac{3}{4\pi^2}\ln\left(b\frac{d}{r_{loc}}\right)} - r_{loc}\right)^4 \qquad (28)$$

where $b$ is a constant, and the weak second order dependences are associated only with $\phi$ and $r_{cage}$. Eq.(28) reflects a core feature of NLE theory: all characteristic lengths and energy scales of the dynamic free energy are inter-related [76].

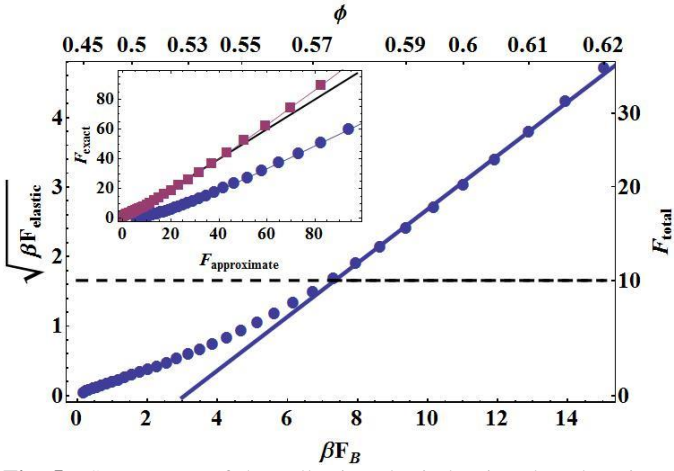

**Fig. 5**: Square root of the collective elastic barrier plotted against the local cage barrier. The right axis shows the total barrier; the upper axis shows the volume fraction. The dashed black line indicates a *total* barrier of $\sim 10 k_B T$ above which $F_{elastic} \propto F_B^2$. Inset: A comparison of the approximate analytic form of the total barrier, Eq. (26), with the numerical result; the black line represents perfect agreement. The blue circles are the results based on Eq. (26) with A = 1, the literal ultra-local limit. The red squares are the result of adjusting A to equal 0.45.

The above results are perhaps reminiscent of phenomenological models such as the Hall-Wolynes (HW) overlapping harmonic well model [87], the landscape equivalent of the shoving model [47], and a recent simulation-inspired generalization [88, 89] involving the transient MSD. However, the HW and shoving models assume an activated event of fixed size, leading to the harmonic form $\beta F_{total} \propto (d/r_{loc})^2$. Our result is qualitatively distinct due to the presence of a local barrier which has a different dependence on the $r_{loc}$ than its collective analog, and the presence of a growing jump length, $\Delta r$, corresponding to the activated event size not being fixed.

### D. Jamming Limit

With increasing volume fraction, the hard sphere fluid eventually jams [90]. Although not accessible in equilibrium, it is of theoretical interest to ask what ECNLE theory predicts in this asymptotic limit. Using the analytic expressions above, this is easily determined with the divergences controlled by a contact value that scales as [77]:

$$g(d) \propto (\phi_{rcp} - \phi)^{-1} \equiv \Delta\phi^{-1} \qquad (29)$$

The predictions as jamming is approached are then:

$$\beta F_B \propto (d/r_{loc})^2 \propto \Delta\phi^{-2} \qquad (30)$$

$$\beta F_{elastic} \propto K_0 \propto G \propto \Delta\phi^{-4} \qquad (31)$$

$$r_B \propto \left(-\ln(\Delta\phi)\right)^{1/2} \qquad (32)$$

$$\Delta r_{eff} \propto -\ln(\Delta\phi) \qquad (33)$$

The localization length vanishes, the length scales associated with hopping diverge logarithmically, and the energy scales diverge as inverse power laws.

## IV. Alpha Relaxation Results

### A. Mean Alpha Time

The generic alpha time follows from Eq.(13) as,

$$\tau_\alpha = \tau_s + \tau_K \approx \tau_s + \tau_{hop} e^{\beta F_{elastic}} \qquad (34)$$

This assumes collective elastic physics only renormalizes the NLE theory barrier, consistent with our view that the elastic fluctuation has a cost and facilitates the local cage scale re-arrangement. Our working expression for the alpha time is thus:

$$\tau_\alpha = \tau_s \left[1 + \frac{2\pi}{\sqrt{K_0 K_B}} e^{\beta(F_B + F_{elastic})}\right] \qquad (35)$$

### B. Comparison to Experiment and Simulation

Recent simulations and hard sphere colloid suspension experiments have documented the failure of ideal MCT critical power law form of the alpha time [91, 92]. In Fig. 6 we present no adjustable parameter calculations, in two representations, based on both ECNLE theory and the prior NLE approach; the vertical scale is shifted to align the theory curve with the data at lower (non-glassy) volume fractions. NLE theory appears to be accurate only when barriers are relatively low, as recently documented irrespective of integral equation theory input [19,93]. On the other hand, including collective elastic effects dramatically improves the predictions at high $\phi$, although the growth still appears to be slightly too weak, possibly due to the PY structural input to the theory employed here. The monodisperse hard sphere theory predicts a quantitatively larger alpha time which seems expected given experiments and simulation employ polydisperse models which speed up dynamics [19].

### C. Numerical Calculations and Comparison to Phenomenological Formulas

Fig. 6 shows alpha time calculations based on Eqs.(4), (5b), (34) and (35) using the total barrier, the local barrier only, and the BCMF short time only. At low volume fractions, the alpha time is essentially equal to the short relaxation time associated with dressed binary collisions. Although a NLE barrier is present at $\phi > \phi_A \approx 0.43$, it is sufficiently low that the relaxation time is still dominated by binary collision physics until $\phi \approx 0.5$. Comparing to Fig.3 shows that for $0.5 < \phi < 0.54$ the activated contribution to relaxation is controlled by the local non-cooperative process, and the elastic barrier is effectively negligible. At $\phi_X \approx 0.54$, the hopping timescale begins to dominate and the *growth* of the elastic barrier becomes appreciable, ultimately being the most important factor controlling the increase of the alpha time beyond $\phi' \approx 0.57 - 0.58$.

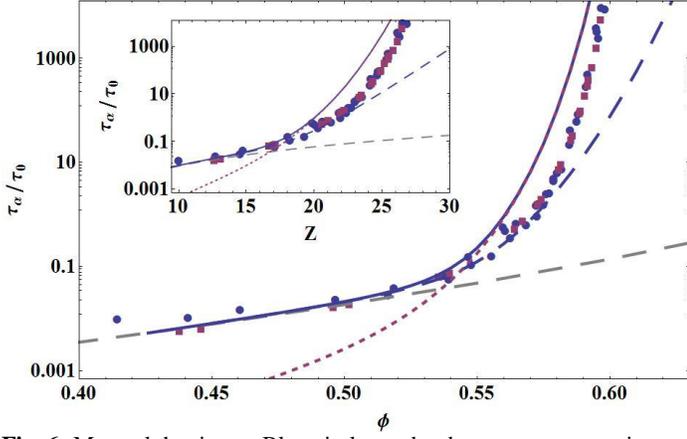

**Fig. 6:** Mean alpha times. Blue circles and red squares are experimental and simulation data for hard sphere suspensions and fluids, respectively [91, 92]. The solid blue curve is the ECNLE result using Eq. 35, the dashed blue curve is the hopping time calculated using only the local barrier in Eq. 12, the gray dashed curve is the short time of Eq. 4. The inset shows the same data and theory plotted against the reduced pressure calculated using Eq. 39. The symbol and line types have the same meaning as in the main figure.

In the log-linear representation of the main frame of Fig.6 there is always significant curvature. Berthier and Witten [94] proposed that the reduced pressure (compressibility factor, $Z = \beta P/\rho$) of a hard sphere fluid is analogous to inverse temperature of a thermal liquid, $Z \leftrightarrow T^{-1}$. The inset to Fig. 6 explores our theoretical predictions in this "Angell–like" representation using the Carnahan-Starling equation of state [70]:

$$Z = \frac{1}{3}\frac{1+2\phi+3\phi^2}{(1-\phi)^2} + \frac{2}{3}\frac{1+\phi+\phi^2}{(1-\phi)^3}. \quad (36)$$

The local NLE theory alpha time displays an interesting behavior in this representation. At low pressures, it is dominated by the dressed binary collision process and yet appears effectively Arrhenius; thus the apparent Arrhenius behavior is a result of both the low barrier local activated process and binary collision physics. At higher pressures, the NLE barrier becomes sufficiently high ($Z \geq 18$) that it controls the relaxation time, and again an apparent Arrhenius behavior is predicted over a rather narrow range of Z but with a *larger* effective non-cooperative barrier. The emergence of super-Arrhenius behavior at high pressures ($Z > 22-23$) is almost entirely due to the collective elastic barrier.

Fig. 7 compares our calculations with phenomenological forms, including Arrhenius behavior

$$\tau_\alpha \sim e^{cZ}, \quad (37)$$

a Vogel-Fulcher-Tamman (VFT) expression (divergence at $Z_0$) [1]:

$$\tau_\alpha \sim e^{\frac{K}{Z_0-Z}}, \quad (38)$$

and the non-singular (below jamming) Bassler or parabolic law [95, 96]:

$$\tau_\alpha \sim e^{CZ^\delta}, \quad \delta = 2 \quad (39)$$

where c, K and C are empirical constants. The VFT formula fits our calculations very well over ~12 orders of magnitude above $\phi' \sim 0.57 - 0.58$. An equally good (or better) fit is afforded by the generalized elastic model of Leporini and coworkers [89, 88]:

$$\ln(\tau_\alpha/\tau_0) = \frac{A}{r_{loc}^2} + \frac{B}{r_{loc}^4} \quad (40)$$

which is nonsingular below jamming where A and B constants.

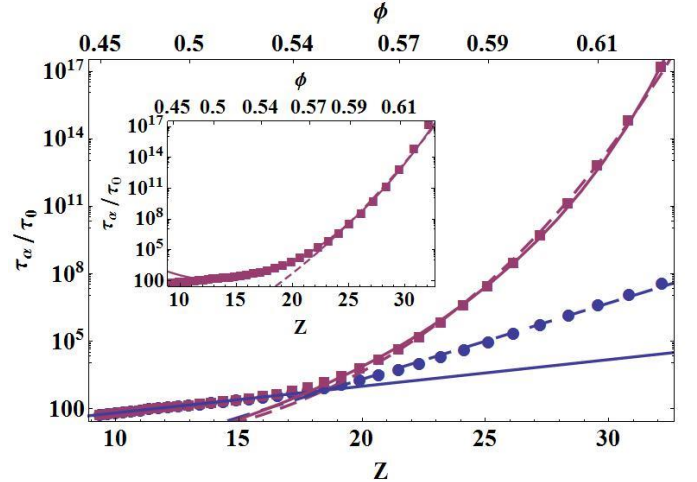

**Fig. 7:** Analysis of the theoretical alpha time as "data". The red squares are the theory results using Eq. 35 and the blue circles are the local hopping time of Eq. 12 using only the local NLE barrier, plotted against the reduced pressure. The hopping time can be fit by two Arrhenius laws, one in the low Z regime(solid blue) and one in the higher Z regime(dashed blue). The red solid curve shows the VFT law of Eq. 37 fit to the full theory for the highly over compressed regime, $Z > 18$. The dashed red curve is a fit of the form proposed by Puosi et.al. [88], Eq. 40, for $Z > 18$. The inset shows the full theory fit to Eq. 38(dashed) and Eq. 41(solid red).

The inset show the Bassler form also works well over 12 orders of magnitude above $\phi'$. Globally, ECNLE theory captures three regimes of relaxation: a low pressure apparent Arrhenius regime, a high pressure super-Arrhenius regime well fit by the Bassler law, and an intermediate smooth crossover regime where the growth of the local and collective barriers are comparable at $\phi' \sim 0.57-0.58$. Our results are qualitatively consistent with simulations[97] that found both non-cooperative and cooperative activated dynamics coexist up to $\phi \approx 0.59$ for the polydisperse hard sphere model studied.

ECNLE theory contains no singularities below jamming, and fits to VFT and the Bassler-like laws are not unique in the sense that the range in Z such forms "work" widens if one empirically increases (decreases) $Z_0$ ($\delta$), consistent with simulation findings. To further emphasize this point, we consider the non-singular expression

$$\ln(\tau_\alpha / \tau_0) = bZ + aZ^2 \qquad (41)$$

where $a$ and $b$ are numbers. This form encodes the two barrier "Arrhenius plus Bassler" viewpoint as simply as possible, in a "parabolic model" spirit [96, 98]. The inset of Fig.7 shows it agrees well with our calculations over 18-19 orders of magnitude.

## V. Comparison to Elastic and Entropy Crisis Models

### A. Shoving Model

The central idea of Dyre's shoving model [49,85] is viscous liquids are solids that flow and the alpha relaxation involves two barriers. An inner compact core (of unknown radius $R_c$) of re-arranging molecules undergoing a (unspecified) local activated event, the realization of which induces a collective barrier associated with shear elasticity and a strain field due to isotropic radial "shoving" of molecules to provide room for the local event. Physical arguments and calculations suggest anisotropic strain field effects are minor. The local volume expansion, $v_c$, is taken to be an empirical, thermodynamic state independent parameter. Ignoring the local barrier, one has:

$$\Delta E_{shove} = v_c G(T) \qquad (42)$$

$$\tau_\alpha = \tau_0 e^{\Delta E_{shove}(T)/k_B T} \qquad (43)$$

where G(T) is the *relaxed* (not infinite frequency) glassy plateau shear modulus which increases with cooling due to anharmonic effects in viscous liquids, and typically $\tau_0 \approx 0.1 \text{ps}$. Whether the barrier associated with the elastic deformation is important or not depends on *both* how large the shear modulus becomes due to emergent rigidity in the deeply supercooled liquid and the absolute magnitude of $v_c$. Until our present work, there has been no physical insight, let alone predictive calculational ability, concerning the latter. Dyre has emphasized there are no finite temperature divergences in the shoving model, and that elastic models have little in common with MCT, dynamic facilitation models, the Adams-Gibbs (AG) model, RFOT or jamming ideas [49].

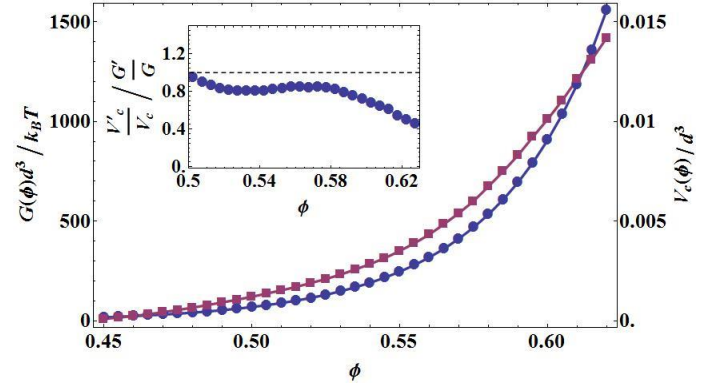

**Fig. 8:** Dimensionless shear modulus (left) and cooperative volume (right) as a function of volume fraction. Inset: Comparison of logarithmic derivatives, showing the relative contribution to the growth of the elastic barrier of the cooperative volume compared with the shear modulus.

Fitting relaxation time data on many materials with Eqs.(42) and (43) yields a $v_c$ typically smaller than the molecular volume [49,99]. The model accounts well for the supra-Arrhenius temperature-dependence over 8-10 orders of magnitude in the viscous liquid regime. At high enough temperatures, experiments find upward deviations from Eq.(43) indicating a larger alpha time than predicted [46,47,49,99]. However, elastic models have been criticized since they have no growing length scale and no explicit mechanism for enhanced cooperativity with cooling [14]. This follows since $v_c$ is presumed constant and the amplitude of the "inner core" expansion that sets the displacement scale of elastic shoving is unknown and assumed independent of temperature or volume fraction. The canonical belief in the glass physics field is that a temperature-dependent barrier implies collective physics and a growing length scale. In RFOT theory the barrier is related to a growing length scale, l, as [14,25]:

$$\Delta E(l^*) = \Delta_0 l^{*\psi} \qquad (44)$$

where $\psi$ is an exponent and $\Delta_0$ sets the energy scale.

We can explore analytically the relation of Eq.(42) with our result for the collective barrier. Using Eq.(22) in Eq.(20), the collective elastic barrier is re-written as

$$F_{elastic} = 12\phi \Delta r_{eff}^2(\phi) r_{cage}^3 K_0 \equiv V_c(\phi) G(\phi) \quad (45)$$

where a "cooperative volume" can be identified as

$$V_c = 20\pi \frac{\Delta r_{eff}^2 r_{cage}^3}{d^2} \quad (46)$$

The shoving-like form follows not by assumption, but as a consequence of the NLE theory prediction $K_0 \propto G$. Most importantly, this "cooperative volume" is predicted to increase with compaction due to an increasing hop distance and cage expansion.

Our result perhaps provides a concrete realization of the speculation of Biroli and Bouchaud [14] that the opposing viewpoints of RFOT and elastic models might be partially bridged if the cooperative volume grows as relaxation slows and $\Delta_0 \propto G$. In terms of Eq.(44), if one identifies $\Delta_0 \propto G$ then the growing length scale $l^* \propto \Delta r_{eff}$, with $\psi = 2$. Its temperature or density dependence reflects the increase of the particle jump length, and involves *both* the localization length and barrier location. Hence, in contrast to the shoving model, the collective barrier increases as a result of *both* a growing length scale and modulus. Fig. 8 and the inset of Fig.3 show in two representations that the cooperative volume is much smaller than a particle volume, which helps resolve the question [14] as to why the length scale $\xi \equiv (6v_c/\pi)^{1/3}$ deduced from fitting data is so small. The main frame of Fig.8 shows the modulus and $V_c$ grow significantly with $\phi$.

It is of high interest to quantitatively establish the *relative* importance of the growing length scale and shear modulus with regards to the increase of the elastically collective barrier. The change in the latter with volume fraction can be analyzed as

$$\beta \frac{\partial F_{elastic}}{\partial \phi} = V_c G \left[ \frac{G'}{G} + \frac{V_c'}{V_c} \right] = \beta F_{elastic} \left[ \frac{d\ln G}{d\phi} + \frac{d\ln V_c}{d\phi} \right] \quad (47)$$

where the prime indicates differentiation with respect to $\phi$. The ratio of the two terms in brackets is the relative growth with densification of $V_c$ compared to $G$. The inset of Fig. 8 shows that over a range of volume fractions (barrier heights) the contributions are similiar, with the modulus contribution ~25% larger. Interestingly, this behavior changes at the same $\phi \approx 0.57 \approx \phi'$ found in section III where $F_{total} \approx 10 k_B T$. In the strongly viscous regime, the modulus growth becomes increasingly more important, and is twice as important as the cooperative volume near kinetic arrest.

In principle, we conclude the pure shoving model idea that the growth of the collective barrier is entirely controlled by the shear modulus does *not* hold in our theory. Figs. 4 and 6 show the collective barrier is unimportant for lower volume fractions, but dominates at high volume fractions where the connection $F_{elastic} \propto F_B^2$ applies (see Fig. 5). In light of the new features associated with a nontrivial cooperative volume, it is not obvious if the shoving model result [85] of $\log(\tau_\alpha/\tau_s) \propto G(\phi)$ agrees with our theory. Remarkably, Fig. 9 shows it works well over ~9-10 decades, as a consequence of the behavior established in the inset of Fig.8. Using $\tau_s \approx 10^{-12}$ in thermal liquids, this range corresponds to $\tau_\alpha > 10^{-8} s$, broadly consistent with experiment [49,99]. For short enough $\tau_\alpha$ the deviation of our alpha time from the shoving model result is upwards, as seen experimentally and in simulation [49,99]. This reflects the impending unimportance of the collective barrier and a smooth crossover to the local cage activation and binary collision regime. This deviation again emerges roughly at $\phi \approx 0.57 = \phi'$.

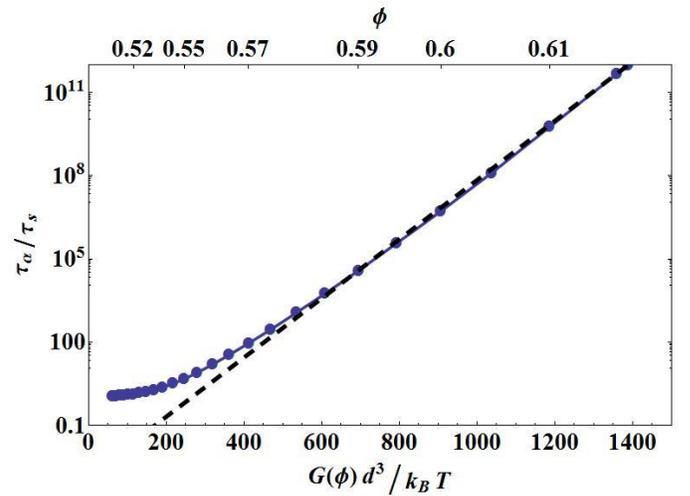

**Fig. 9:** Alpha relaxation time plotted against the dimensionless shear modulus. The smooth curve through the points is a guide to the eye, and the dashed line is the pure shoving model form.

**B. Entropy Crisis Theories**

The classic Adams-Gibbs model [54] is formulated in terms of activated dynamics. At high temperature an Arrhenius process is assumed dominant (barrier $E_A$) reflecting "uncorrelated single particle hopping". At lower temperatures, AG argue hopping becomes correlated and involves z(T) molecules in a "configurational re-arranging region" (CCR), yielding a net alpha time of:

$$\frac{\tau_\alpha}{\tau_0} = e^{z(T) E_A / k_B T} \quad (48)$$

As a further postulate, z(T) is argued to scale inversely with configurational entropy :

$$z(T) \propto (k_B / S_{conf}(T)) \qquad (49)$$

Super-Arrhenius relaxation arises from a decrease of $S_{conf}(T)$ with cooling. No microscopic description of the activated event is provided. Experiments on molecular liquids typically find a small z ~ 3-5 at $T_g$, which contradicts a large CCR picture [1,7,30,100,101].

The basis of an apparent Arrhenius barrier in NLE theory is cage scale dynamics. As a consequence of its 2-barrier nature, ECNLE theory can be written in the AG form,

$$\frac{\tau_\alpha}{\tau_0} \propto e^{\beta(F_B + F_{elastic})} = e^{zF_B/k_BT}, \quad z \equiv 1 + \frac{F_{elastic}}{F_B} \qquad (50)$$

where the theoretically well-defined "cooperativity parameter" grows monotonically upon densification; Fig. 3 suggests z~3 at vitrification. Alternatively, one can follow empirical experimental analyses and use an *apparent* Arrhenius barrier to define the elementary process; this leads (not shown) to modestly larger cooperativity parameters.

The more sophisticated RFOT [23] differs from the AG model in several respects, and high temperature activated process plays no direct role. Recently, Rabochiy and Lubchenko [102] performed an elaborate calculation that combined RFOT with *equilibrium* elasticity ideas for the surface tension. Their result for the mean alpha time is

$$\tau_\alpha \approx \tau_0 \exp\left[cd^3 \frac{B + 4G/3}{TS_{conf}}\right] \qquad (51)$$

where c is a complicated factor related to liquid structure. In essence, the surface tension (numerator of the barrier) is now related to the bulk and shear moduli. The temperature-dependence of the barrier is thus stronger than the AG and pure RFOT expressions. For deeply supercooled liquids the above result is not of elastic model nor ECNLE theory form. Moreover, in elastic and ECNLE theories it is the relaxed glassy plateau modulus that enters, and G is directly correlated with the dynamic localization length, which is not an "equilibrium" elastic property and has a different thermodynamic state dependence.

Another conceptual difference between entropy crisis ideas and our work is that the Kauzman transition (even if it exists) does not lead to literal arrest. Rather, beyond this volume fraction (believed to be below jamming [21]) the system condenses into a sub-extensive number of packing structures which results in a crossover of the equation-of-state (and compressibility, S(q), etc) to a non-liquid-like "free volume" form [19,21,103]. The barriers remain finite, though exceedingly high and inaccessible in equilibrium.

## VI. Nonexponential Relaxation and Dynamical Length Scales

We have focused on the mean alpha time, $\tau_\alpha$. Of course there is a distribution of times and a litany of dynamic heterogeneity (DH) phenomena [6,104]. Whether the latter is kinetic, structural, and/or small scale thermodynamic origin remains debated, as is whether DH fundamentally drives the alpha process or is a generic second order consequence of activated dynamics. There are many propositions of growing dynamic and static length scales, but their inter-relationship, relevance to observables, and relation to a "cooperativity length" is largely unclear. Here we briefly comment on non-exponential relaxation and growing length scales in ECNLE theory.

### A. Stretched Relaxation

Stretched exponential relaxation is ubiquitous, though its interpretation is often clouded by the influence of unresolved secondary or beta relaxations and/or a dependence on the specific correlation function measured. For systems where the alpha process is well resolved, many materials show stretching with an exponent of $\beta_K \approx 0.5$, which at most weakly changes with temperature or density [105, 106]. Significantly, the latter includes the molecular liquids where the strongest translation-viscosity (or rotation) decoupling has been observed [105-108]. These trends are strongly buttressed by recent measurements by Rossler et.al. [28,35] on many molecular liquids which find alpha relaxation stretching is nearly independent of thermodynamic state with $\beta_K \approx 0.5 - 0.6$. Similar behavior is found in simulations of hard sphere fluids where diffusion-relaxation decoupling at the single particle level occurs with increasing volume fraction but with no change in $\beta_K$ [109]; this behavior is well predicted by hard sphere NLE theory [13].

The above observations suggest a minimalist, almost trivial, interpretation in the ECNLE framework. The basic relaxation event is barrier hopping. If this process obeys a Poisson distribution, corresponding to a relaxation time distribution of $P(\tau) = (\tau/\tau_\alpha^2)e^{-\tau/\tau_\alpha}$, then a generic exponential time correlation function becomes [18]:

$$C(t) = \int_0^\infty d\tau P(\tau) e^{-t/\tau} = \frac{2t}{\tau_\alpha} K_2\left(2\sqrt{\frac{t}{\tau_\alpha}}\right) \propto e^{-\sqrt{9t/4\tau_\alpha}} \qquad (52)$$

where $K_2$ is the modified Bessel function of the 2nd kind, and the final form applies over the last half of the decay. This seems to be the simplest possible *purely* dynamical idea for the origin of stretched relaxation with a thermodynamic-state-independent exponent of roughly one half. We are *not* claiming it is a unique or definitive explanation.

### B. Dynamic Length Scales

One possible growing scale in ECNLE theory is the cooperative volume, $V_c$, of Eq.(46). Calculations of $V_c$ as a function of $\log(\tau_\alpha)$ in the inset of Fig.3 show it increases more quickly for shorter relaxation times than in the strongly viscous regime, and the latter growth is slow and roughly logarithmic, increasing by only a factor of ~3 over a 10-12 orders of magnitude change in $\tau_\alpha$. One can ask if these trends are *qualitatively* sensible compared with recent diverse experimental attempts to extract [3,4,110-112] the number of cooperatively moving (or dynamically correlated) molecules, $N_{corr}$. Key experimental findings include a two regime form of $N_{corr}$ versus $\log(\tau_\alpha)$, with faster growth at shorter alpha times, a slower and roughly *logarithmic* growth by a factor of ~3-4 in the deeply supercooled regime, and a rather weak material dependence of $N_c$ on $\log(\tau_\alpha)$. Although the precise connection between $V_c$ and $N_{corr}$ is unclear, it seems natural they would be correlated, consistent with the similarities between the experimental trends and Fig. 3.

Given the long range strain field surrounding the core of the activated event, how to define a growing length scale is not obvious. An energetic criterion is to ask at what distance 90% of the *collective* barrier is attained, and the answer is ~13-15 particle diameters, independent of thermodynamic state. Alternatively, local measures are the jump length, $\Delta r$, or its effective analog, $\Delta r_{eff}$, shown in Fig.3.

### VII. Summary

We have generalized the microscopic force-level NLE theory of cage scale activated single particle hopping to include collective effects associated with the long range elastic distortion required to accommodate local re-arrangement. The elementary relaxation event is of a mixed local-nonlocal spatial character, in contrast to diverse theories based on compact domains. For simplicity, the ideas have been developed and implemented in the context of hard sphere fluids. The alpha relaxation process involves two distinct, but inter-related (primarily via the transient localization length), local and collective barriers. The local true "Arrhenius" process of NLE theory is sub-dominant at all volume fractions except in a relatively narrow intermediate volume fraction crossover region. At lower volume fractions, it is obscured by (dressed) binary collision physics, while at high volume fractions it is coupled with, and ultimately dominated by, collective elastic effects. The super-Arrhenius collective barrier is more strongly dependent on volume fraction, dominates the highly viscous regime, and is well described by the Bassler, VFT, or parabolic model forms. The collective barrier growth with volume fraction is determined by both the shear modulus (or inverse square of the localization length) and jump length, with the former increasing more rapidly with $\phi$ than the latter.

The theory predicts slow dynamics and structure are related at two different levels. The cage scale barrier computed using NLE theory depends on microscopic details and local packing. However, the collective barrier required to facilitate the cage scale re-arrangement controls the deeply supercooled regime, and is determined by longer range elasticity and continuum-like particle displacements. Hence, ECNLE theory seems consistent with the argument that in the deeply supercooled regime a structure/dynamics connection cannot be fully made at the local single particle level [113].

Four theoretically well-defined dynamic crossovers are predicted. (i) A barrier and transient localization emerges (naïve MCT transition) at $\phi_A \sim 0.432$. (ii) Activated motion becomes relevant in the sense of the hopping times exceeding the non-activated process time scale ($\tau_{hop} > \tau_s$) at $\phi_X \sim 0.53$ where $F_{tot} \sim 3k_BT$. (iii) The rate of change of the collective barrier exceeds its local analog at $\phi' \approx 0.57 - 0.58$ where $F_{tot} \sim 10k_BT$. This correlates with smooth crossovers in many dynamical properties such as the bending up (down) of the alpha time (cooperative volume) as a function of volume fraction, and also the local and collective barriers beginning to obey $F_{elastic} \propto F_B^2$. (iv) The collective barrier exceeds its local analog close to a kinetic vitrification at $F_{tot} \sim 30 \, k_BT$.

Concerning the five distinguishing aspects of glassy dynamics theories stated in the Introduction [5], ECNLE theory is characterized by: (i) no divergences above zero Kelvin or below jamming, (ii) connections to thermodynamics are a consequence of dynamics, (iii) no literal nor avoided critical points, (iv) at zeroth order dynamic heterogeneity is a consequence of activated motion, and (v) a specific molecular-level physical picture underlies the activated relaxation process. A corollary to (i) is that the theory *does* predict a divergent relaxation time and length scale ($\Delta r$, or cooperative volume) as the jammed state is approached and a contact network emerges.

The present work provides the foundation for constructing a theory for thermal liquids based on mapping to a reference hard sphere fluid, as described in the following paper II, where limitations and possible generalizations and theoretical improvements of the approach are also discussed.

**Acknowledgements**. We acknowledge support from the Division of Materials Science and Engineering, U.S. Department of Energy, Office of Basic Energy Sciences via ORNL. Discussions and/or correspondence with Jeppe Dyre, Ryan Jadrich, Mark Ediger, Alexei Sokolov, Dino Leporini and Ernst Rossler are gratefully acknowledged.

**Appendix A: Cage Expansion Amplitude**

The average cage expansion length scale of Eq.(16) was obtained by pre-averaging the outward particle motion over all angles. This is motivated by, and consistent with, the scalar isotropic nature of NLE theory. Here an alternative derivation of mean cage expansion is presented which incorporates directional motion.

The cage center is taken to be the coordinate system origin. Consider a particle at position $\vec{z}$ which hops in the *vector* direction $\overrightarrow{\Delta r}$. After such a motion, the distance of the particle from the cage center is $r_{cage} + r_{exp} = \sqrt{(\vec{z} + \overrightarrow{\Delta r})^2} = \sqrt{z^2 + \Delta r^2 + 2z\Delta r \cos\theta}$, where $\theta$ is the angle between the vectors $\vec{z}$ and $\overrightarrow{\Delta r}$. Taking $r_{exp} \to 0$ defines the maximum angle, $\psi$, for which the cage must expand in order to accommodate the motion, where $\cos\psi = \frac{r_{cage}^2 - z^2 - \Delta r^2}{2z\Delta r}$. Performing the angular average over the hopping direction (taking $\vec{z}$ to lie on the z axis) we then obtain

$$\langle r_{exp} \rangle_\theta = \frac{2\pi}{4\pi} \int_0^\psi d\theta \sin\theta\, r_{exp}(\theta) \quad (A1)$$
$$= \frac{1}{2} \int_0^\psi d\theta \sin\theta \left(\sqrt{z^2 + \Delta r^2 + 2z\Delta r \cos\theta} - r_{cage}\right)$$
$$= \frac{1}{12z\Delta r}\left[r_{cage}^3 + 2(z+\Delta r)^3 - 3r_{cage}(z+\Delta r)^2\right]$$

The mean cage expansion follows from averaging $\langle r_{exp} \rangle_\theta$ over the cage volume

$$\Delta r_{eff} = \frac{\int d\Omega \int_{r_{cage}-\Delta r}^{r_{cage}} dz\, z^2 \langle r_{exp}\rangle_\theta}{\int d\Omega \int_0^{r_{cage}} dz\, z^2} \quad (A2)$$
$$= \frac{3}{r_{cage}^3}\left[\frac{r_{cage}^2 \Delta r^2}{12} + \frac{r_{cage}\Delta r^3}{48} - \frac{\Delta r^4}{120}\right]$$

As in the main text, taking $\Delta r \ll r_{cage}$ yields the leading order result

$$\Delta r_{eff} \to \frac{3}{12}\frac{\Delta r^2}{r_{cage}} = \frac{1}{4}\frac{(r_B - r_{loc})^2}{r_{cage}} \quad (A3)$$

Eq.(A3) is qualitatively identical to Eq.(17), differing only by a numerical prefactor of 8/3. We believe the relation $\Delta r_{eff} \propto \Delta r^2$ is robust. In the big picture, one might view the numerical prefactor as the single adjustable quantity for hard spheres.

---